\documentclass[aps,pra,twocolumn,amsmath,amssymb]{revtex4-2}
\usepackage{graphicx}
\usepackage{caption}
\usepackage{subcaption}
\usepackage{dcolumn}
\usepackage{bm}
\usepackage{orcidlink}
\usepackage{hyperref}
\hypersetup{
    colorlinks=true,
    linkcolor=red,
    citecolor=green,
    urlcolor=blue}

\begin{document}
\title{Self-Trapping and Liouvillian Spectral Transitions in a Dissipative Bose–Hubbard Dimer}
\author{Subhendu Saha \orcidlink{0000-0003-1660-9358}}
\email{ssphy\_rs@caluniv.ac.in}
\author{Raka Dasgupta \orcidlink{0000-0003-2148-4641}}
\email{rdphy@caluniv.ac.in}
\affiliation{Department of Physics, University of Calcutta, $92$ A. P. C. Road, Kolkata $700009$, India}
\date{\today}

\begin{abstract}
We study the self-trapping transition in dissipative double wells using mean-field theory and exact diagonalization. We find that, similar to the conventional interaction-driven transition from oscillatory dynamics to self-trapping, dissipation alone can also induce self-trapping-like localization in one well. The resulting state, although transient and decaying, persists for a substantial time.  At the mean-field level, the interaction- and dissipation-driven transitions are discontinuous and continuous, respectively. When interaction and dissipation both are present, their interplay produces three dynamical regimes: (i) at weak interaction, increasing dissipation drives a transition from oscillations to self-trapping; (ii) at moderate interaction, self-trapping at weak dissipation, oscillatory dynamics at intermediate dissipation, and re-entrant self-trapping at strong dissipation; and (iii) at strong interaction, self-trapping persists but is weakened by dissipation. We connect the transition with the Liouvillian spectrum. It is observed that purely dissipation-driven transition coincides with an exceptional point in the spectra, and a peak in the Liouvillian gap, marking the onset of the quantum-Zeno regime.  In the presence of interactions, however, the onset of the Zeno regime is shifted to dissipation strengths larger than that required for the self-trapping transition.
\end{abstract}
\maketitle


\section{Introduction}\label{Sec.1}
Ultracold bosons in  double-well potentials are often considered as Bose–Josephson junctions (BJJ). Such systems not only exhibit Josephson oscillations analogous to those observed in conventional Josephson junctions, but also display a variety of intriguing dynamical phenomena arising from their intrinsic nonlinearity.  The discovery of self-trapping transition in ultracold bosons trapped in a double well is one such example. It was theoretically predicted~\cite{Smerzi1997, Raghavan1999} that beyond a critical value of the interatomic interaction strength, the particles tend to occupy one particular well only, thus breaking the underlying $\mathcal{Z}_2$ symmetry. This phenomenon was subsequently observed experimentally in a bosonic Josephson junction formed by two coupled $\mathrm{Rb}^{87}$ condensates, where the transition from Josephson oscillations to macroscopic self-trapping was directly demonstrated~\cite{Albiez2005}.

While the mean-field picture provides an intuitive description of the transition, a finite condensate is essentially a quantum many-body system. The two-mode Bose–Hubbard model therefore provides a natural microscopic framework for investigating the self-trapping phenomenon, which allows quantum fluctuations~\cite{Milburn1997}, finite-particle-number effects~\cite{Julia2010}, and deviations from mean-field dynamics~\cite{Buonsante2010}. Other studies have shown that quantum fluctuations modify the mean-field picture and can lead to long-lived or quasi-self-trapped dynamics in finite systems~\cite{Zhou2024, Soares2025, Bardin2026}. More recent investigations have extended the concept of self-trapping to dynamical tunneling and nonlinear dynamics in modulated, anharmonic potentials~\cite{wuster2025}. These developments establish self-trapping as an important probe for studying quantum many-body dynamics in two-mode systems. 

The extension of these two-mode condensates to dissipative Bose–Hubbard dimers~\cite{Secli2021} offers a test bed for investigating how particle loss and environmental coupling modify the transition and associated dynamical regimes. This is extremely important from experimental perspectives, because in realistic condensates, the two-mode system is generally not perfectly isolated. Dissipation in ultracold atomic systems is not restricted to an uncontrolled coupling to the environment; it can also be engineered and spatially localized. In particular, a localized dissipative potential can be implemented by selectively removing atoms from a confined region of a Bose–Einstein condensate, as demonstrated experimentally~\cite{Barontini2013, Gericke2008, Takafumi2017}. Theoretical studies have also explicitly considered Bose–Hubbard systems with particle losses localized at a single lattice site, establishing a microscopic framework for describing the competition between coherent tunneling and local dissipation~\cite{Kepesidis2012, Brazhnyi2009}. The competition between interactions and dissipation in such systems has been extensively investigated using mean-field and exact quantum approaches, revealing effects such as dissipation-induced coherence~\cite{Witthaut2008} and modified tunneling dynamics~\cite{Witthaut2009}.  More generally, dissipative Bose-Hubbard systems provide a versatile platform for investigating non-equilibrium many-body phenomena and related emergent behaviors that have no direct counterpart in isolated systems~\cite{Kordas2015epjs, kordas2015ann}. The dissipative dynamics here is most often modeled in the framework of Lindblad Master equation \cite{Breuer2007,Lindblad1976}. The comparison between mean-field approaches and exact quantum evolution is particularly important in this setting, since dissipation modifies not only the population dynamics but also quantum correlations and fluctuations~\cite{Witthaut2011}. The distinction between mean-field dynamics and the exact quantum evolution of a finite number of particles is particularly important in the presence of dissipation, since particle loss changes both the particle-number distribution and the coherence between the two modes. Studies of open Bose-Hubbard systems have shown that mean-field descriptions can capture the qualitative dynamics, while quantum corrections become relevant for finite particle numbers and in regimes where fluctuations are enhanced~\cite{Witthaut2009, kordas2015ann, Witthaut2011}. 

An efficient way of characterizing dissipative dynamics is provided by the spectrum of the Liouvillian superoperator. Unlike a Hamiltonian, whose eigenvalues are real for a closed quantum system, Liouvillian eigenvalues are generally complex. These eigenvalues determine the characteristic relaxation and oscillation modes of the density matrix, with the real parts governing decay rates and the imaginary parts determining coherent oscillation frequencies~\cite{Mori2023, Mori2023rev,Mori2024,Haga2021,Haga2024}. The eigenvalue with zero real part corresponds to the stationary state, while the eigenvalue closest to the imaginary axis determines the slowest asymptotic relaxation process~\cite{Minganti2018, Casteels2017,Kessler2012,Fink2018,Vicentini2018,Marko2015}. The associated Liouvillian spectral gap therefore provides a natural measure of the relaxation timescale of the dissipative condensate. 

\begin{figure}
    \includegraphics[width=\linewidth]{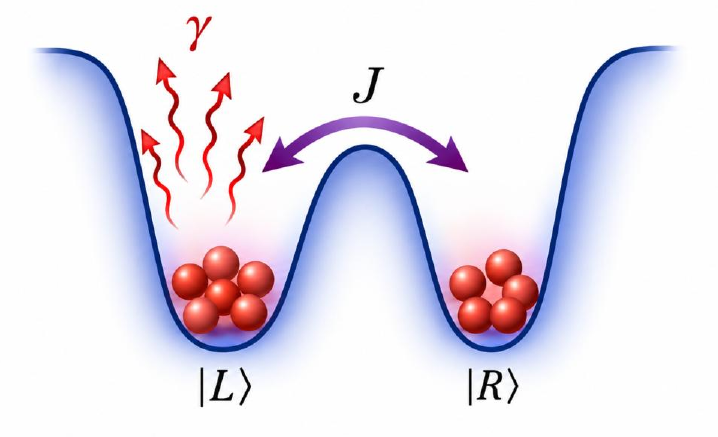}
    \caption{Schematic diagram of a dissipative bosonic Josephson junction. Bosons trapped in a double-well potential undergo coherent tunneling between the wells, while localized particle loss with rate $\gamma$ acts on Well 1.}
    \label{bjj}
\end{figure}

In this work, we study the dissipative dynamics of a two-mode Bose-Einstein condensate (as in Fig. \ref{bjj}) and investigate the time-dependent population dynamics using mean-field and exact diagonalization approaches. We focus on the competition between interparticle interactions and dissipation, with a particular emphasis on the transition between the oscillatory and self-trapped regimes. We identify them as two distinct phases, and compute the relevant order-parameter and the corresponding susceptibility. Additionally, we examine the real and imaginary parts of the leading nonzero Liouvillian eigenvalues and demonstrate how the spectral properties provide a dynamical characterization of the transition between these two regimes. The Liouvillian gap is further used to quantify the characteristic relaxation timescale, and its dependence on the interaction and dissipation strengths is investigated. 

The rest of this work is organized as follows: In Sec.\ref{Sec.2} we present the model Hamiltonian for double-well system and an outline for solving it using Lindblad master equation. Sec.\ref{Sec.3} comprises of population imbalance dynamics using mean-field and exact diagonalization methods. In Sec.\ref{Sec.4} we compute the order parameter and susceptibility for characterizing the interaction-driven and dissipation-driven transitions. Sec.\ref{Sec.5} is devoted to the Liouvillian spectrum analysis. Finally, we conclude our results in Sec.\ref{Sec.6}, and outline possible future prospects.


\section{Theoretical Setup}\label{Sec.2}
We begin with the simplest Bose-Josephson Hamiltonian \cite{Gati2007, Shchesnovich2010} describing BEC in a symmetric double-well potential $V(\mathbf{r})$, given by 
\begin{equation}
H = -J(a_L^\dagger a_R + a_R^\dagger a_L) + U_L n_L^2 + U_R n_R^2
\end{equation}
where $a_j^\dagger (a_j), j = L,R$ are bosonic creation (annihilation) operators for the wave functions $\varphi_{L,R}$ localized at the potential minima of the left and right well respectively. Also, $n_j$($a_j^\dagger a_j$) is the corresponding number operator, $J$ is the single atom tunneling rate, $U_{L,R} = g/2 \int d^3 \mathbf{r} \varphi_{L,R}^4$ is the on-site interaction strength between bosons, where $g=4\pi \hbar^2 a_s /m$, $a_s$ being the s-wave scattering length and $m$ is the atomic mass. In a trap having two quasidegenerate lowest energy levels, one can assume $U_L \approx U_R = U$ with a good accuracy.
In a situation where atoms are engineered to dissipate from one of the wells (say, \emph{left well}) with a dissipation coefficient $\Gamma$, one can describe the dynamics using the Lindblad Master Equation \cite{Breuer2007,Lindblad1976,Gorini1976}
\begin{equation}\label{lindblad}
	\dot{\rho} = -\frac{i}{\hbar}[H,\rho] + \Gamma \left(a_L \rho a_L^\dagger - \frac{1}{2}\{a_L^\dagger a_L, \rho\}\right)
\end{equation}
Here $\rho$ is the density matrix of the system. The first term in Eq.(\ref{lindblad}) governs the coherent dynamics in the system and the dissipation part is responsible for gradual loss of coherence between the two wells of the double well trap. The interplay of the on-site interaction $U$ and dissipation strength $\Gamma$ leads to interesting dynamical regimes described by the master equation. 

One can explicitly solve Eq.(\ref{lindblad}) for non-interacting $(U=0)$ case (experimentally, this amounts to make $a_s$ negligible using the Feshbach resonance). In case of interacting atoms, Eq.(\ref{lindblad}) cannot be solved exactly. Thus, one has to apply mean-field (MF) approximation to arrive at the following dynamical equations:
\begin{equation}\label{dyneqn}
    \begin{aligned}
        \dot{z} &= -\sqrt{1-z^2} \sin \phi - \frac{\gamma}{4}(1-z^2) \\
        \dot{\phi} &= \frac{z}{1-z^2} \cos \phi + \Lambda n z - \frac{\gamma}{4} \tan \phi \\
        \dot{n} &= -\frac{\gamma}{4} n (1+z)
    \end{aligned}   
\end{equation}
where $\Lambda = U N/J$ is the reduced atomic interaction in BEC and $\gamma = \Gamma \hbar/J$ is the normalized atom removal rate \cite{Smerzi1997, Raghavan1999}. The mean-field variables $z, \phi$ and $n$ correspond to the average quantities:
\begin{equation}
    z = \frac{\langle n_L\rangle - \langle n_R\rangle}{\langle n_L\rangle + \langle n_R\rangle}, \quad e^{i\phi} = \frac{\langle a_L^\dagger a_R \rangle}{\sqrt{\langle n_L\rangle \langle n_R\rangle}}, \quad n = \frac{\langle n_L\rangle + \langle n_R\rangle}{N}
\end{equation}
where $z$ is the population imbalance, $\phi$ is the relative phase, and $n$ is the total number of particles at any instant. Throughout this paper, we scale all energies by $\hbar$, thus converting energy scales to angular frequency scales. Further, the unit of energy is set by $J=1$. In the mean-field analysis, we absorb the initial particle number $N$ into the interaction strength.  Accordingly, the interaction parameter $U$ is to be interpreted as the effective nonlinear interaction scale, rather than as the microscopic two-particle interaction for a single boson. In this scale, $\Lambda$ depends only on this effective interaction $U$.

We also use exact diagonalization (ED) as a complementary approach to our mean-field calculations. For a total number of \(N\) bosons distributed between the left and right wells, the basis in the absence of dissipation is
$$
|n_L,n_R\rangle=|N,0\rangle,|N-1,1\rangle,\ldots,|0,N\rangle,
$$
with $n_L+n_R=N$. The two-site Bose-Hubbard Hamiltonian is then represented by an $(N+1)\times(N+1)$ matrix. In the presence of dissipation, the total particle number is no longer conserved. We therefore include all states satisfying $n_L+n_R\leq N$. The resulting Hilbert-space dimension is
$$
D=\sum_{n=0}^{N}(n+1)=\frac{(N+1)(N+2)}{2}.
$$
We can express Eq.(\ref{lindblad}) as 
$$
\dot{\rho}=\mathcal{L}[\rho],
$$
where \(\mathcal{L}\) denotes the Liouvillian superoperator, represented by a \(D^2\times D^2\) matrix, that can be written in following form:
\begin{equation}
\begin{aligned}
     \mathcal{L} &= -i(H\otimes I-I \otimes H^T
    ) + \gamma a_L \otimes a_L^* \\ &-\frac{\gamma}{2} \left[a_L^\dagger a_L \otimes I + I \otimes (a_L^\dagger a_L)^T \right]
\end{aligned}
\label{Liouvillian}
\end{equation}
The Liouvillian superoperator is constructed in the truncated Fock-space basis and numerically diagonalized using exact diagonalization (ED). Eq.(\ref{lindblad}) can be solved with the help of Eq.(\ref{Liouvillian}) to obtain the time evolution of $\rho$ and extract the population dynamics from it. One can also solve the Liouvillian eigenvalue equation $\mathcal{L}|R_\alpha\rangle = \lambda_\alpha |R_\alpha\rangle$ and find the Liouvillian spectrum consist of complex eigenvalues $\lambda_\alpha = \mathrm{Re}[\lambda_\alpha]+ i \mathrm{Im}[\lambda_\alpha]$ as a function of $\gamma$. This eigenvalue spectrum is further used to calculate the Liouvillian gap 
 \begin{equation}\label{gap}
    \Delta_L = - \max\limits_{\lambda_\alpha \neq 0} \mathrm{Re}[\lambda_\alpha]
\end{equation}
and characterize the dissipative dynamics of the system.
 

\section{Self-trapping}\label{Sec.3}
It is well-established that self-trapping can be implemented by increasing the on-site interaction and keeping the system totally isolated from the environment. An important question that follows is how robust this behavior remains in the presence of dissipation. Since realistic systems are inevitably coupled to their environment, dissipation can significantly alter the coherent tunneling and interaction-induced localization. That is why we consider a constant dissipation (strength $\gamma$) from one of the wells in the presence of a small finite interaction and find that surprisingly, dissipation, too, can independently aid in inducing self-trapping.
 
Before investigating the joint effects of having interaction and dissipation together, we separately chart out two distinct ways of achieving self-trapping in the BJJ system: (i) interaction-driven (the conventional self-trapping in a closed system), and (ii) dissipation-driven (in absence of on-site interaction). The two types of self-trapping transitions are then compared by studying the corresponding order parameters and susceptibilities. 
\subsection{Interaction-driven transition }
\begin{figure}
    \includegraphics[width=\linewidth]{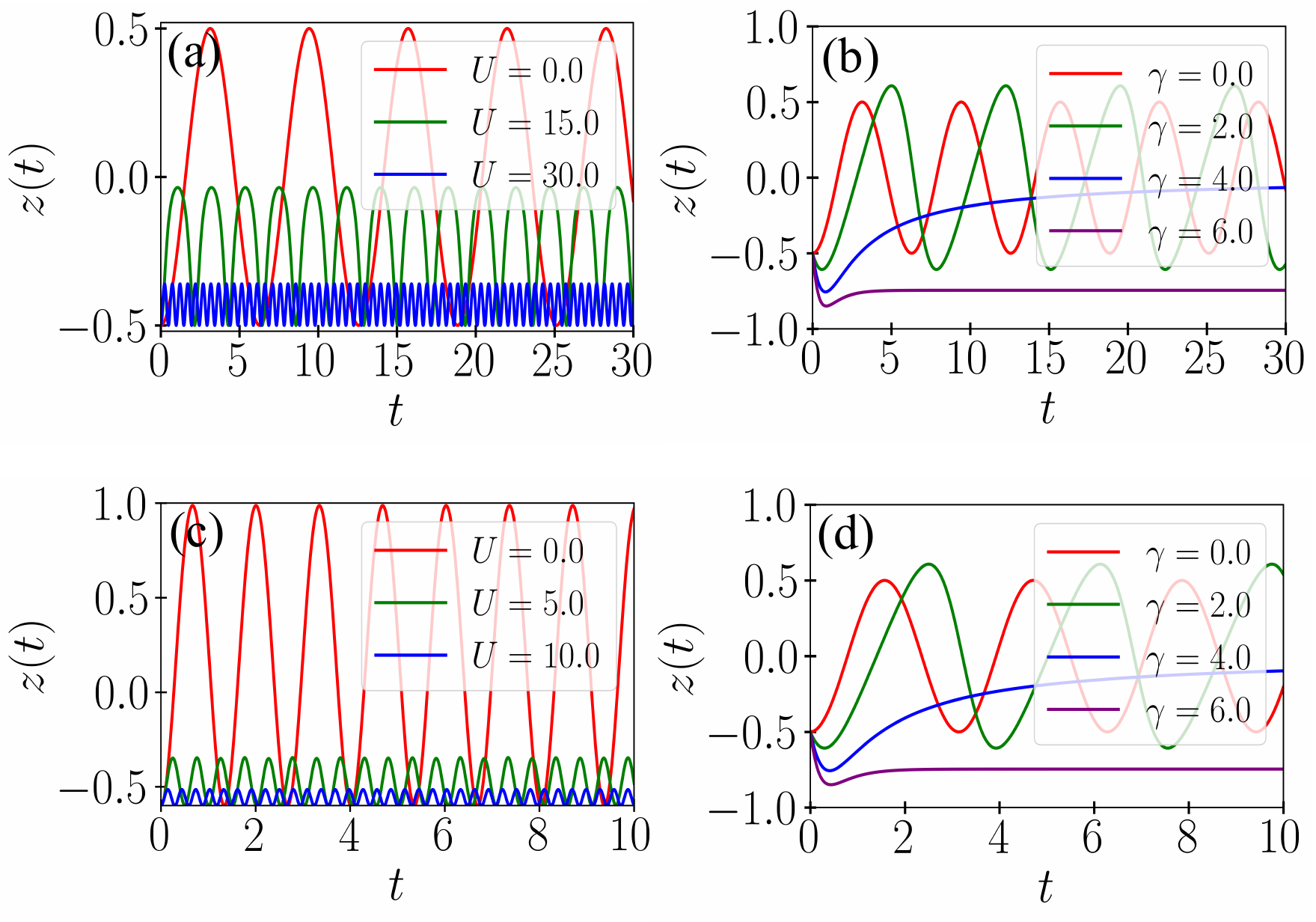}
    \caption{Plots of population imbalance $z(t)$ for interaction and dissipation-driven dynamics, respectively. (a) MF results varying $U$ at $\gamma=0$ (b) MF results varying $\gamma$ at $U=0$, (c) ED results varying $U$ at $\gamma=0$ for $N=5$ (d) ED results varying $\gamma$ at $U=0$ for $N=4$}
    \label{pop_imb}
\end{figure}
We first study the population imbalance as a function of time for interaction-driven case using both MF and ED method, keeping $\gamma=0$, i.e., no particle is dissipating from either of the wells. We find that, in the non-interacting limit $(U=0)$, the MF picture shows large-amplitude oscillations of imbalance between the two modes, indicating regular population exchange driven by the hopping dynamics. Oscillations remain nearly periodic over the entire time window considered. This is the expected Rabi-like regime \cite{Smerzi1997} where interaction do not restrict the particle transfer. It is evident from Fig.\ref{pop_imb}(a) that the dynamical characteristics change noticeably as the interaction strength $U$ is increased further. After crossing a certain $U$ value, the oscillation amplitude is strongly reduced compared to the non-interacting regime, while the oscillation frequency becomes higher. This happens because the imbalance no longer explores the full phase space accessible and instead remains confined to a narrower region. This suggests that interactions restrict the transfer of particles between the two modes, termed as macroscopic self-trapping in literature \cite{Smerzi1997, Raghavan1999, Albiez2005}.

We obtain nearly similar results in Fig.\ref{pop_imb}(c) using ED method, where periodic oscillations around $z=0$ suggests that the average imbalance is close to zero. In contrast to the $U=0$ case, where the populations repeatedly exchange between the two modes, the system  remains localized near a preferred initial imbalance configuration throughout the evolution in the presence of finite interaction. Thus, we found that increasing the interaction strength progressively suppresses the coherent population transfer and confines the system to a final imbalance state.
\subsection{Dissipation-driven transition}
\begin{figure}
    \includegraphics[width=\linewidth]{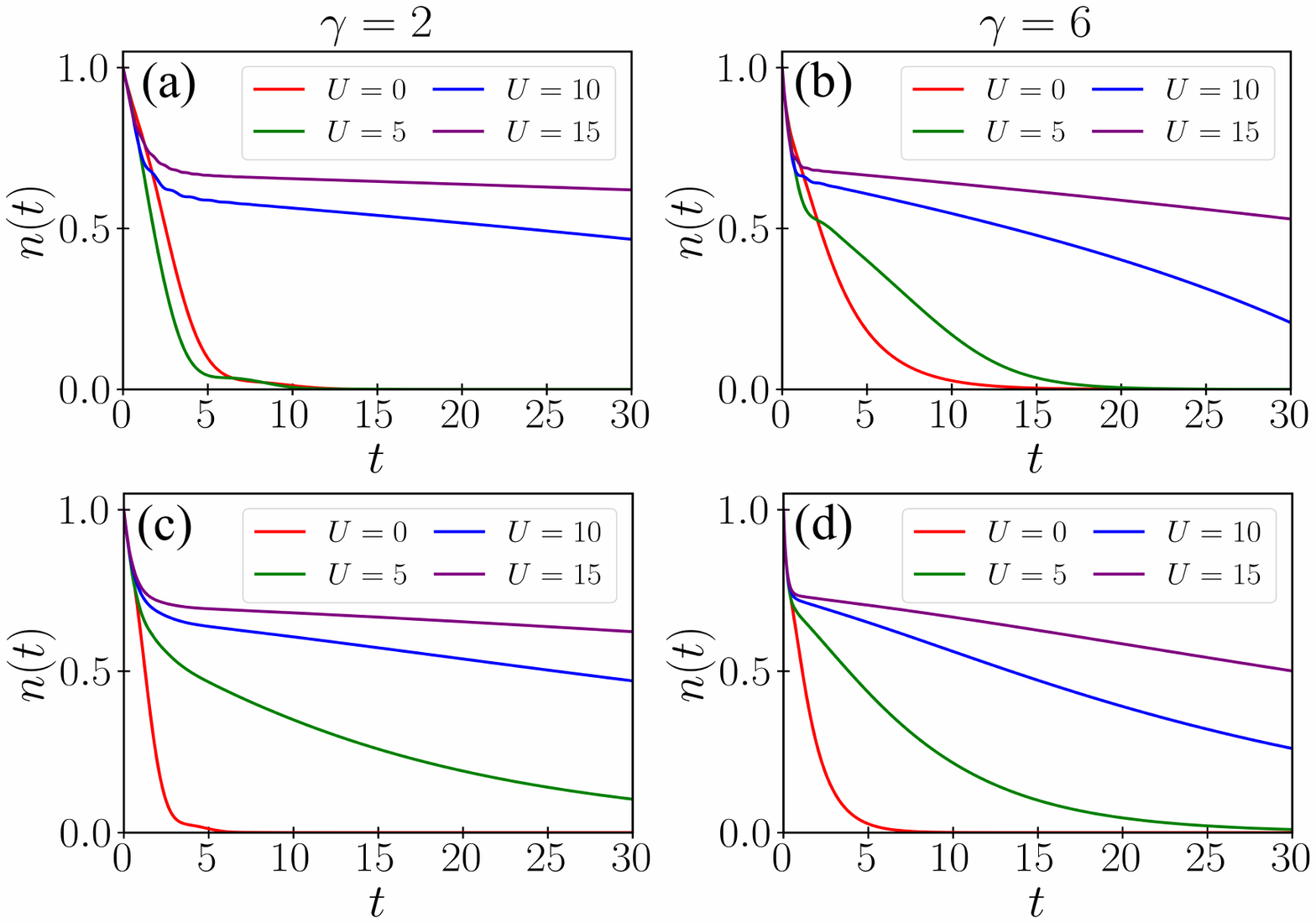}
    \caption{Plots of time evolution of the normalized particle number $n(t)$. Panels (a)-(b) shows MF method and (c)-(d) shows ED method.}
    \label{num}
\end{figure}
To address the effect of dissipation, we next examine the time evolution of the population imbalance for different values of the dissipation strength $\gamma$ in Fig.\ref{pop_imb}(b) using MF method. As we observed earlier, in the absence of dissipation $(\gamma=0)$, the imbalance exhibits persistent oscillations with a nearly constant amplitude, reflecting the coherent nature of the underlying dynamics. Introducing a weak dissipation channel $(\gamma=2)$ does not immediately destroy this oscillatory behavior. Instead, the system continues to display pronounced oscillations over the entire observation window, indicating that coherent dynamics remain dominant over dissipative effects in this regime. Increasing $\gamma$ further gradually damps the transient oscillatory dynamics observed at early times and the system enters a non-oscillatory state. In the oscillatory regime, $z$ switches between positive and negative values, while in the non-oscillatory regime $z$ is always negative, and eventually reaches a steady value. This behavior implies that once dissipation is sufficiently strong, particle loss from the left well dominates over coherent inter-well transfer, effectively confining all remaining particles in the dissipation-less right well. We identify $\gamma_c \approx 4$ as the critical dissipation strength at which the system crosses from oscillatory population-transfer dynamics to a non-oscillatory, dissipation-induced self-trapped regime in the non-interacting limit $(U=0)$. Below $\gamma_c$, coherent tunneling sustains oscillations of the population imbalance, whereas above $\gamma_c$, localized particle loss suppresses coherent population exchange and drives the system toward a strongly imbalanced transient state.

This phenomenon bears resemblance to the interaction-induced self-trapping. There are two major differences though: (i) In the interaction-driven self-trapped state, $z$ shows oscillations around a mean-value. In the dissipation-driven case, $z$ reaches a nearly steady value after initial fluctuations. (ii) In the interaction-driven model, the system is closed, i.e., the total number of particles remain conserved. In the dissipation-driven model, particles keep escaping from the well: evolving slowly towards a stationary configuration with zero particles. The long-time dynamics here is fully governed by dissipative relaxation rather than coherent population exchange. 

The studies using ED (as shown in Fig.\ref{pop_imb}(d)), too, confirm a similar behavior. The value of $\gamma_c$ for the self-trapping transition is also nearly same for both ED and MF. 


\subsection{Nature of transition: interaction-driven vs dissipation-driven}
\begin{figure}
    \includegraphics[width=\linewidth]{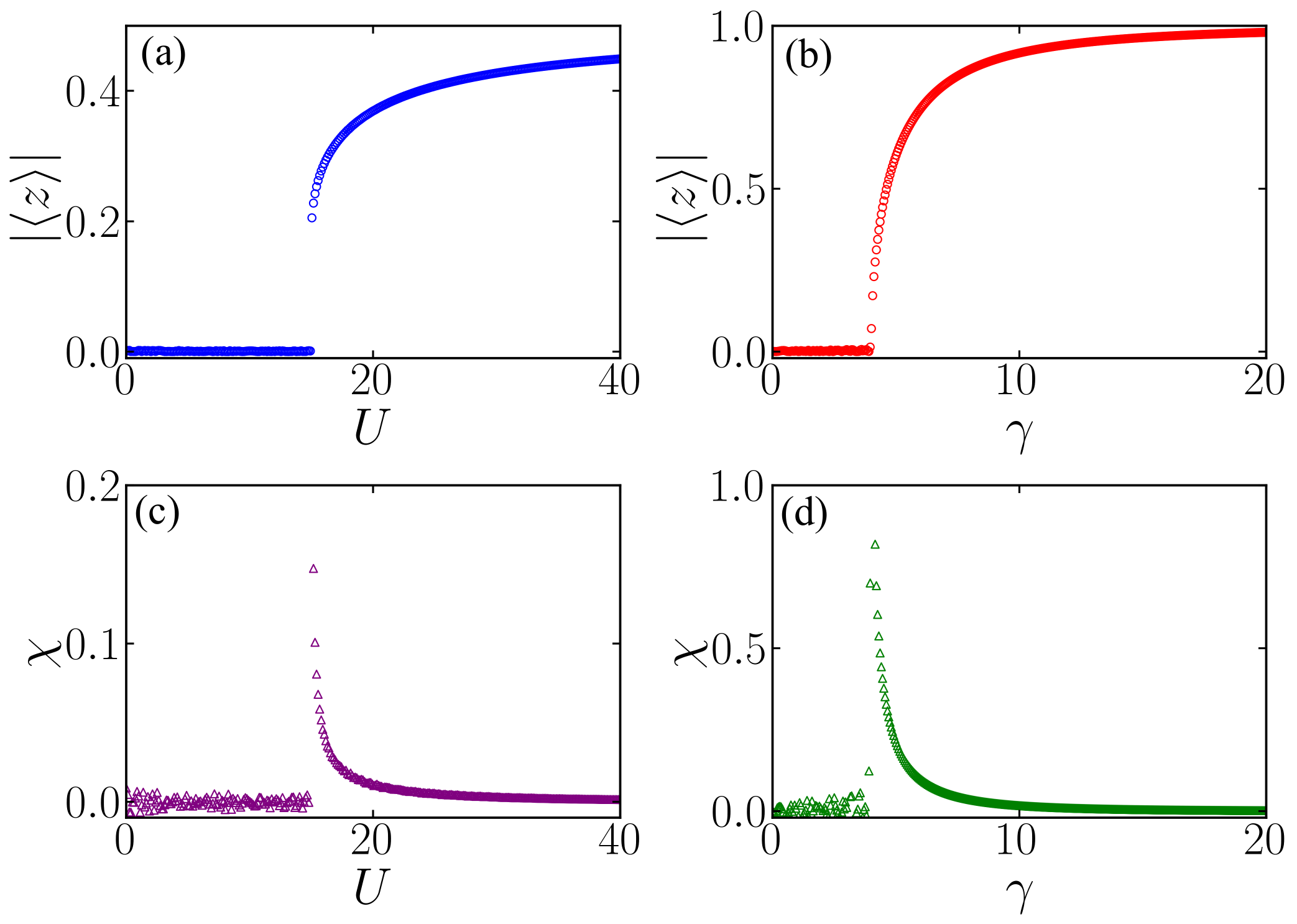}
    \caption{Plots of order parameter $|\langle z \rangle|$ and susceptibility $\chi$ as functions of the interaction strength $U$ and dissipation rate $\gamma$. Panels (a) and (c) show the results as $U$ is varied at $\gamma=0$, while (b) and (d) correspond to varying $\gamma$ at $U=0$. Here $J=1$, $N=1$ were used throughout.}
    \label{op}
\end{figure}
One of the major difference between interaction-driven self-trapping and dissipation-driven self-trapping is that, the former results in a  self-trapped steady state, while the later leads to a  self-trapped transient state: eventually culminating in  an empty double-well in the long-time limit. From Fig.\ref{num}(a)-(d), it is evident that some particles can survive till $t>30$ if the nonlinearity is strong enough, and at least till $t\approx 5$ in the absence of $U$. Typical experimental values (assuming, for example, $\mathrm{Rb}^{87}$ condensates, 1064 nm lattice, lattice depth $V_0=8 E_R$, $E_R$ being the recoil energy ) lead to $J\approx 0.5\times10^3 \mbox{s}^{-1}$ and unit of time $t\approx 2$ ms. Thus, the transient state can exist till $10$ ms, or even longer: a time-frame long enough for quantum observations and measurements \cite{Streed2006, Barontini2013, Labouvie2016, Tian2025}. Since the lifetime of the self-trapped state is comparable to, or longer than, the intrinsic dynamical timescale of the underlying cold-atom system, the state can provide a sufficiently long-lived platform for quantum-state manipulation.

To distinguish further between the oscillatory and self-trapped states driven by dissipation, we compute the absolute time averaged population imbalance, ie. $|\langle z\rangle|$ as order parameter and corresponding susceptibility $\chi$ as functions of the interaction parameter $U$ and dissipation strength $\gamma$. We do this time average over the range of $t=0$ to $t=50$, which is well sufficient for realistic experiments. Fig.\ref{op}(a) depicts a pure interaction-driven transition where the order parameter remains zero up-to a critical $U$, and then performs a discontinuous jump. On the other hand, as $\gamma$ is varied in a purely dissipative transition in absence of $U$, the order parameter shows a continuous evolution. Thus, we can classify interaction-driven and dissipation-driven non-equilibrium transitions as 1st order (discontinuous) and 2nd order (continuous) transitions respectively. We proceed further to investigate the corresponding susceptibility for both cases. It is found that the susceptibility blows up in a much sharper way in case of the $\gamma$-driven transition, again signaling a continuous phase transition. We also note that both the transitions in the order parameter and susceptibility happens at the same critical dissipation strength $\gamma_c$, identified from the population dynamics previously. Thus, the dynamical crossover is consistently reflected in the response of the order parameter.
\section{Interplay of Interaction and Dissipation}\label{Sec.4}
\begin{figure}
    \includegraphics[width=\linewidth]{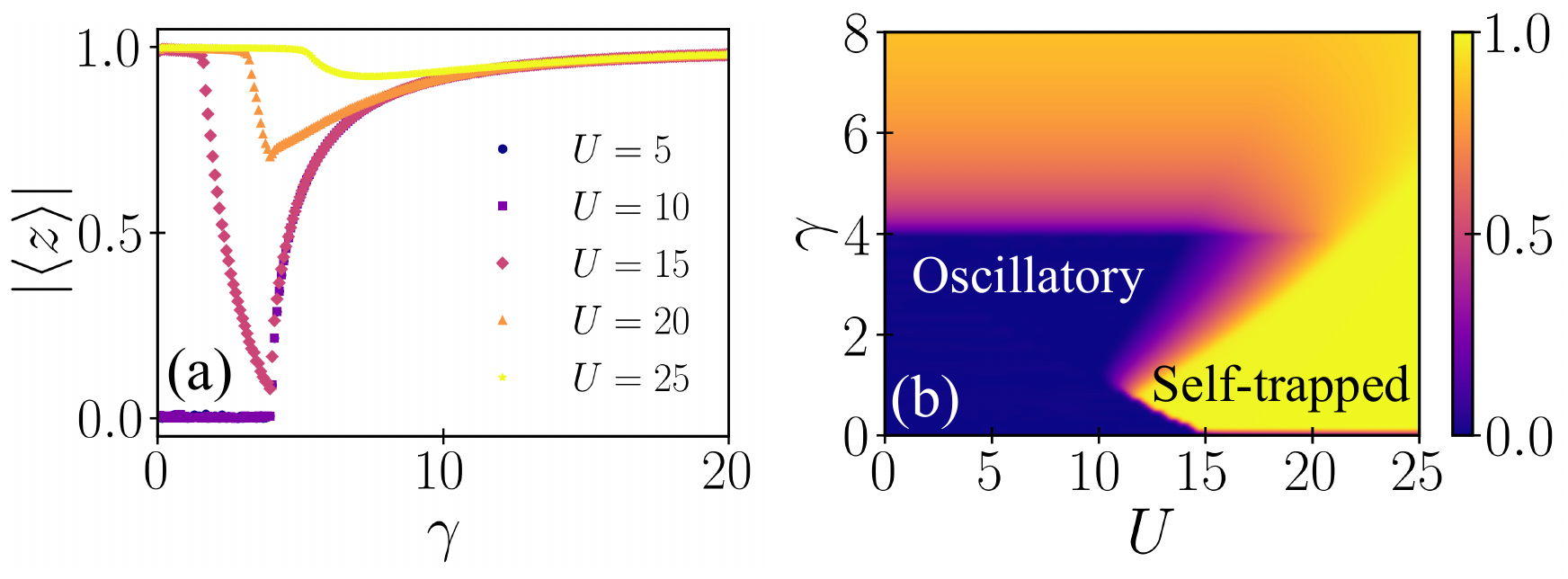}
    \caption{Dissipation-driven crossover between oscillatory and self-trapped dynamics. (a) Order parameter $|\langle z\rangle|$ as a function of $\gamma$ for different $U$ values. (b) Phase diagram in the $U-\gamma$ parameter plane, showing the oscillatory and self-trapped dynamical regimes. The color scale represents $|\langle z\rangle|$, with larger values indicating stronger self-trapping.}
    \label{phsdiag}
\end{figure}
Next we investigate the collective effects of interaction and dissipation, when both are present simultaneously. Fig.\ref{phsdiag}(a) portrays the dependence of the order parameter on the dissipation strength $\gamma$ and interaction strength $U$. For weak interaction, the system predominantly remains in the oscillatory regime over a substantial range of $\gamma$. In particular, for $U=5$ or $U=10$, the time-averaged imbalance remains close to zero till $\gamma \leq 4$. This indicates that the coherent tunneling dynamics is sufficiently strong to prevent the development of a persistent population imbalance. The effect of dissipation becomes increasingly important in larger $\gamma$, where the imbalance increases and the system eventually enters into a self-trapped regime. Thus, for weak interactions, self-trapping is not present over the entire dissipation range and requires sufficiently strong dissipation. 

A qualitatively different behavior emerges at intermediate interaction strength. For $U=15$, the system initially exhibits a strongly self-trapped state with weak dissipation, characterized by $|\langle z\rangle|$ close to unity. However, increasing $\gamma$ leads to a pronounced suppression of the imbalance. Around an intermediate dissipation strength, the order parameter becomes very small, signaling a crossover from the small initial self-trapped regime to an oscillatory regime. Upon further increasing the dissipation, the imbalance increases again and approaches a large value. Consequently, the system displays a re-entrant sequence, as $\gamma$ is increased. This non-monotonic response demonstrates that dissipation does not act merely as a mechanism that monotonically suppresses or enhances localization. Instead, its effect depends crucially on the relative strength of interaction and tunneling. 

The same competition is manifested for stronger interactions, $U=20$ and $U=25$. For weak dissipation, both systems exhibit a large population imbalance, reflecting the strong tendency towards interaction-induced self-trapping. For $U=20$, increasing $\gamma$ produces a pronounced reduction of $|\langle z\rangle|$, followed by a recovery at stronger dissipation. In contrast, for $U=25$, it remains close to unity over most of the parameter range and shows only a relatively weak suppression at intermediate dissipation. Therefore, increasing the interaction strength progressively stabilizes the self-trapped state against the dissipative perturbation.

The overall behavior is clearly illustrated by the phase diagram in Fig.\ref{phsdiag}(b), where the color scale represents the magnitude of the long-time averaged imbalance. The low-$|\langle z\rangle|$ region forms the oscillatory phase, while the high-$|\langle z\rangle|$ region corresponds to self-trapped dynamics. The boundary separating these two regimes is strongly dependent on both $U$ and $\gamma$, demonstrating that neither interaction nor dissipation alone determines the dynamical state. At small $U$, the oscillatory region extends up to the same critical dissipation strength $\gamma_c$. As $U$ is increased, the self-trapped region expands, indicating the increasing dominance of interaction-induced localization. The phase boundary also develops a pronounced curvature, consistent with the non-monotonic behavior observed in Fig.\ref{phsdiag}(a). In particular, the re-entrant behavior at intermediate interaction strengths demonstrates that $\gamma_c$ cannot be treated as a universal threshold independent of $U$.  We therefore distinguish the critical dissipation $\gamma_c$ in the non-interacting limit from the interaction-dependent boundary separating oscillatory and self-trapped dynamical regimes.

The phase diagram further reveals that the oscillatory regime is most prominent at intermediate values of both $U$ and $\gamma$. This region corresponds to the parameter range in which dissipation is sufficiently strong to destabilize the interaction-induced self-trapped dynamics, but not yet strong enough to produce the large imbalance observed at stronger dissipation. The recovery of the self-trapped regime at large $\gamma$ therefore gives rise to a re-entrant structure in the dynamical phase diagram.

These results can be understood in terms of the competition between coherent tunneling, interaction-induced dynamics, and dissipation. Josephson tunneling favors the exchange of particles between the two wells and therefore promotes oscillatory behavior. Increasing $U$, on the other hand, enhances the nonlinear energy cost associated with population re-distribution and favors a persistent population imbalance. Dissipation modifies this competition by changing the long-time evolution of the system. In the intermediate-dissipation regime, it can suppress the imbalance and restore oscillatory dynamics, whereas at stronger dissipation the long-time dynamics again develops a substantial imbalance. The resulting competition produces the non-monotonic dependence of the dynamical order parameter on $\gamma$.

An important consequence is that the dissipative dynamics cannot be interpreted simply in terms of a monotonic loss of coherence or suppression/enhancement of self-trapping. Instead, the response is intrinsically nonlinear: dissipation can first destabilize an interaction-induced self-trapped state and subsequently restore a large population imbalance at stronger coupling. The increasingly robust self-trapping observed with increasing $U$ further demonstrates that interaction provides a stabilizing mechanism against the dissipation-induced oscillatory regime.


\section{Liouvillian Spectrum}\label{Sec.5}
\begin{figure}
    \includegraphics[width=\linewidth]{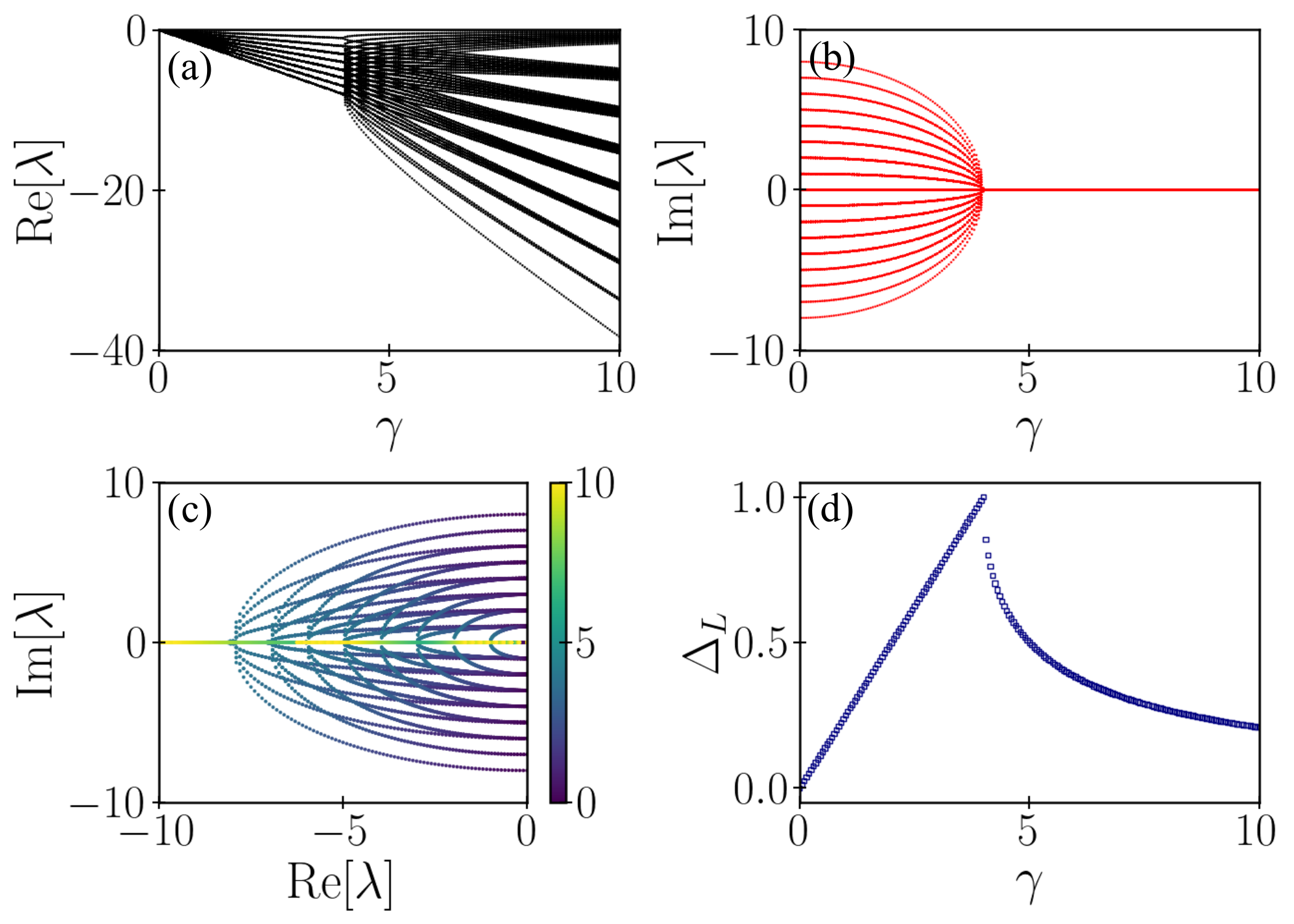}
    \caption{Evolution of the Liouvillian spectrum with $\gamma$ at $U=0$. (a) Real and (b) imaginary parts of the Liouvillian eigenvalues as functions of $\gamma$. (c) Trajectories of the Liouvillian eigenvalues in the complex plane as $\gamma$ is varied, the color indicating the dissipation strength. (d) Liouvillian gap ($\Delta_L$) as a function of $\gamma$, showing a maximum at $\gamma_c$ followed by a decrease for larger $\gamma$.}
    \label{lv}
\end{figure}

A useful way to understand the change in dynamical behavior is to examine the the Liouvillian spectrum as the dissipation strength $\gamma$ is varied \cite{Nakagawa2021}. The complex eigenvalues contain information about both the decay and the oscillation of the dynamical modes. In particular, the real part of the Liouvillian eigenvalue determines its decay rate, whereas its imaginary part is associated with the oscillation frequency. Fig.\ref{lv} shows how these eigenvalues evolve with $\gamma$, together with the corresponding spectral gap $\Delta_L$ given by Eq.(\ref{gap}). 
At weak dissipation, the eigenvalues are characterized by finite imaginary parts, while their real parts remain very small (exactly zero at $\gamma=0$). As $\gamma$ increases, the real part branches out as shown in Fig.\ref{lv}(a). The imaginary eigenvalues, as shown in Fig.\ref{lv}(b), appear as pairs symmetrically distributed around $\lambda=0$. 

The situation changes progressively as $\gamma$ increases. The branches in the imaginary part shown in Fig.\ref{lv}(b) collapse at a critical $\gamma$ value ($\gamma_c$) along the $\lambda=0$ line. Thus, $\gamma=\gamma_c$ is the exceptional point of this spectrum. At the same time, at $\gamma_c$, the real parts of the eigenvalues begin to spread towards negative values, as seen in Fig.\ref{lv}(a). In other words, the dominant modes retain a pronounced oscillatory character up to $\gamma_c$, though the characteristic frequencies associated with the oscillatory modes decrease with increasing $\gamma$. There is a clear change in the structure of the Liouvillian spectrum, as the imaginary eigenvalues drop to zero at $\gamma_c$. 

The spectral behavior becomes clearer in the complex eigenvalue plane shown in Fig.\ref{lv}(c). At small $\gamma$, the eigenvalues extend predominantly along the imaginary direction, indicating that the spectrum is controlled by modes with appreciable oscillatory frequencies. As $\gamma$ increases, these eigenvalues move in a negative-real direction. The trajectories gradually become compressed towards the real axis, and near $\gamma_c$ the conjugate branches approach each other. For larger $\gamma$, the spectrum is dominated by eigenvalues with small imaginary components and increasingly negative real parts. Consequently, the movement of the eigenvalues away from the imaginary axis with increasing $\gamma$ reflects dissipative relaxation. At the same time, the collapse of the imaginary components indicates suppression of the Josephson-like oscillations. The spectral evolution therefore provides a simple interpretation of how increasing dissipation changes the character of the dynamical modes. 

The Liouvillian gap ($\Delta_L$), shown in Fig.\ref{lv}(d), provides another way to track this spectral restructuring as well as the underlying dynamics. Starting from a value close to zero at $\gamma=0$, $\Delta_L$ increases linearly as $\gamma \rightarrow 0$. Here, relatively weak dissipation leads to very low coupling to the environment. So, the system dynamics is mainly dominated by the coherent Hamiltonian, and it retains a significant contribution from the oscillatory modes. The environment acts as a weak perturbation to the system, so increasing $\gamma$ simply speeds up the relaxation process, taking less time to reach the steady state. As $\gamma$ increases, $\Delta_L$ reaches a maximum value at $\gamma_c$, where the imaginary branches of the Liouvillian spectrum have nearly collapsed onto the real axis. This represents the maximum relaxation rate of the system associated with the spectral modes governing the dynamics. Moreover, this point marks the transition or crossover where the coherent energy scales of the Hamiltonian perfectly balance out the environmental dissipation. At this specific value of $\gamma_c$, the system reaches its steady state faster than at any other dissipation. Beyond this, $\Delta_L$ continuously decreases as $\gamma$ increases, resulting in a non-monotonic dependence of $\Delta_L$ on the dissipation strength. This is the regime of the quantum Zeno Effect \cite{Zhu2014, Nakagawa2020, Daley2014, Popkov2018}. 
\begin{figure}
    \includegraphics[width=0.85\linewidth]{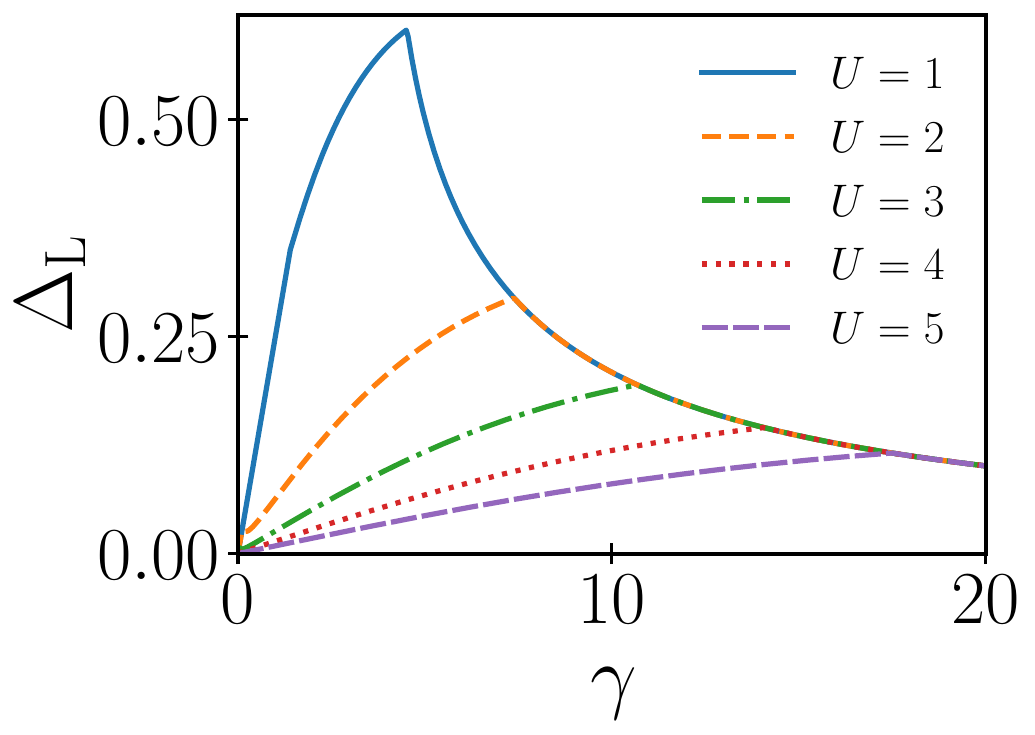}
    \caption{$\Delta_L$ as a function of $\gamma$ for different $U$ values. The shift of the corresponding peak value of $\Delta_L$ with increasing $U$ characterize the dependence of the dissipative relaxation spectrum on the interaction strength.}
    \label{lvu}
\end{figure}\\\
When dissipation is very strong, the interaction with the environment acts as a continuous measurement that effectively freezes the system's coherent transitions. Here, population transfers are suppressed, and relaxation to the global steady state slows down dramatically, causing the gap to close. 

An important feature of these results is that the strongest spectral change occurs at an intermediate value of $\gamma$, rather than at the largest dissipation considered. This is reflected in the maximum of $\Delta_L$ near $\gamma_c$. For much larger $\gamma$, the eigenvalues continue to move deeper into the negative-real region, but the corresponding value of $\Delta_L$ decreases. Hence, increasing dissipation beyond this characteristic region does not simply amplify the spectral response. Instead, once the oscillatory branches have collapsed, further dissipation mainly modifies the decay rates of the already non-oscillatory modes. 

In absence of $U$, the peak of $\Delta_L$ corresponds to the same $\gamma_c$ where the transition from oscillatory to self-trapping occurs, meaning, the system relaxes fastest at this point. This corresponds to a consistent physical picture: at  weak dissipation, relaxation is slower because the dynamics is predominantly oscillatory, and at strong dissipation, relaxation is again suppressed by the quantum Zeno effect.  Consequently, between these two regimes, i.e., exactly at the oscillatory-to-self-trapped transition point, the Liouvillian gap reaches its maximum, corresponding to the fastest relaxation.
\begin{figure}
    \includegraphics[width=\linewidth]{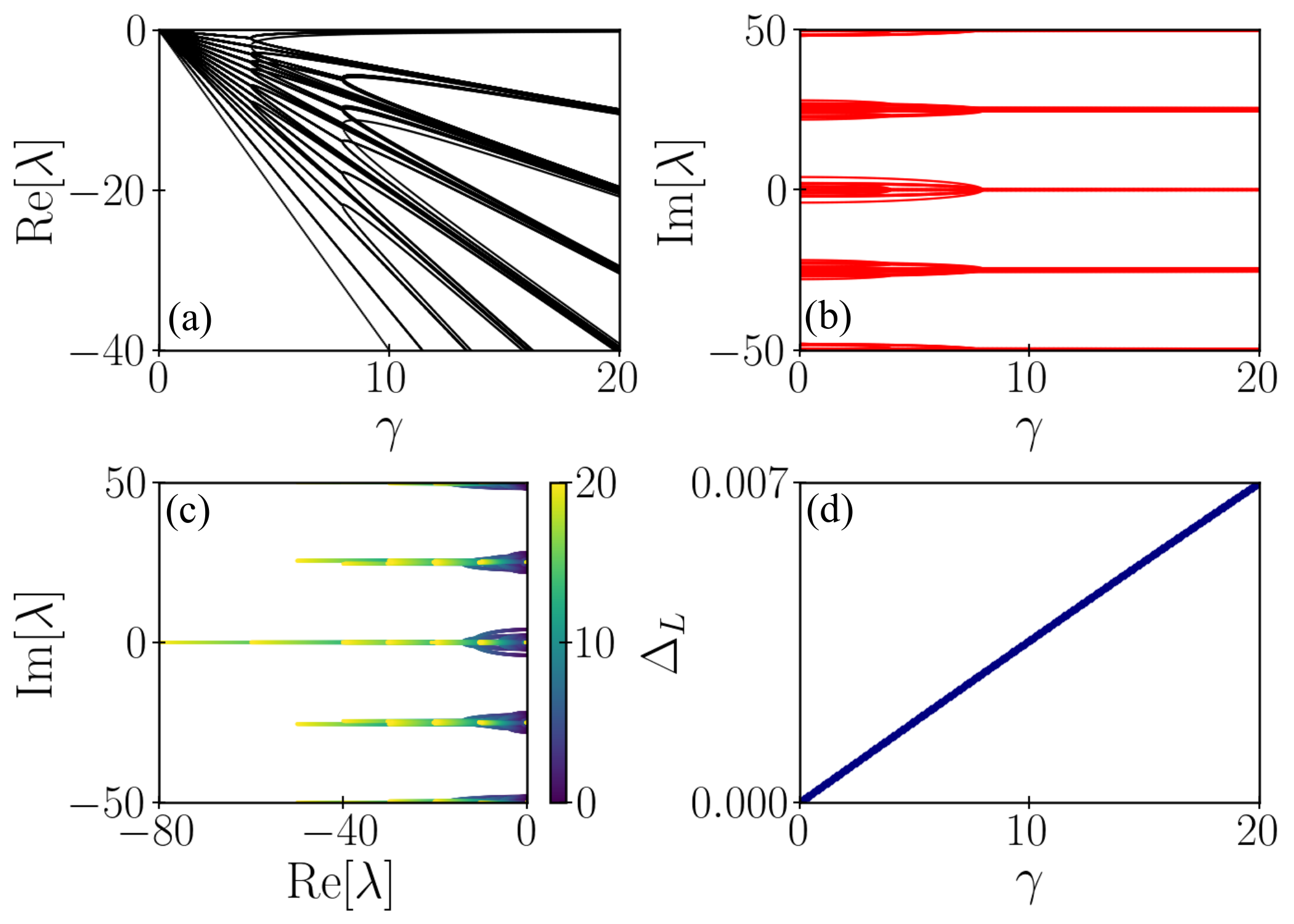}
    \caption{Evolution of the Liouvillian spectrum with $\gamma$ at $U=25$. (a) Real and (b) imaginary parts of the Liouvillian eigenvalues as functions of $\gamma$. (c) Trajectories of the Liouvillian eigenvalues in the complex plane as $\gamma$ is varied, the color indicating the dissipation strength. (d) Liouvillian gap ($\Delta_L$) as a function of $\gamma$. The gap becomes very small at $U=25$.}
    \label{lv2}
\end{figure}
To illustrate the effect of finite interaction, we again plot $\Delta_L$ for several $U$ values in Fig.\ref{lvu}. We can see that increasing $U$ alters the position and shape of the peak. With increasing $U$, the peak shifts to higher $\gamma$ values, and becomes less sharp in nature. Introduction of a non-zero $U$ turns on an additional energy scale in the dynamics. As a result, there is fragmentation in the imaginary spectra, as shown in Fig.\ref{lv2}. At this point, the system does not have an exceptional point anymore: though there is still an overall crossover from oscillatory decaying state to self-trapped decaying state. The point of fastest relaxation, too, shifts towards right: indicating that a higher dissipation value is now required in order to achieve the fastest relaxation. This happens because strong interaction pushes the energy scales of the coherent Hamiltonian upward. It requires a much stronger $\gamma$ to overcome these robust interaction energies and enter into the localized Quantum Zeno regime \cite{Li2026}. The exact correspondence between the self-trapping transition and the position of the Liouvillian gap peak is thus, lost in the presence of $U>0$. 


\section{Conclusion}\label{Sec.6}

In this work, we have investigated the emergence of self-trapping in a dissipative Bose–Hubbard dimer. We have examined how coherent tunneling, on-site interaction, and particle loss compete to determine the long-time dynamics of the system. By combining mean-field analysis with exact diagonalization, we have shown that self-trapping can arise through two distinct mechanisms: (i) via nonlinear interactions and (ii) via dissipation from one of the wells. The comparison between the complementary MF and ED approaches also demonstrates that, although finer details like oscillation amplitude, relaxation timescale etc. might differ from MF to ED, the qualitative trends remain exactly similar. Even the value of the critical $\gamma_c$ for initiating self-trapping in a non-interacting system of particles are same as calculated from MF and ED. Since MF is valid for larger number of particles (as it neglects fluctuations), and our ED accounts for the full quantum solution but for a very small number of particles: we conclude that our key results are very much robust and does not depend on the system size. 

For interaction-driven dynamics, increasing the interaction strength suppresses coherent population transfer and drives the system from an oscillatory regime toward a self-trapped state. In contrast, particle loss from one of the wells provides an independent route to localization. The resulting phase diagram reveals that the effect of dissipation is strongly non-monotonic. In particular, for intermediate interaction strengths, increasing the dissipation can first destroy interaction-induced self-trapping, restore oscillatory dynamics, and then produce self-trapping again at stronger dissipation. Thus, dissipation cannot be regarded simply as a mechanism that suppresses coherent dynamics; its effect depends sensitively on the relative strength of interaction and hopping.

We use population imbalance as an order parameter to distinguish between the statistical nature of interaction-driven and dissipation-driven transitions.  Within the mean-field description, the interaction-driven transition is accompanied by a discontinuous change of the order parameter, whereas the dissipation-driven transition develops continuously. This is also corroborated by the behavior of the susceptibility $\chi$, that diverges in the $\gamma$-driven transition in a much sharper way. This difference provides a useful dynamical characterization of the two routes to self-trapping, and emphasizes that the presence of environmental coupling can qualitatively modify the nature of the transition.

Finally, we move into Liouvillian description of this open quantum system for looking at its dynamical features from a different perspective. The complex eigenvalue spectrum of the Liouvillian superoperator gives a comprehensive representation of the dynamical modes. In the absence of particle-particle interaction,  the imaginary parts of the leading Liouvillian eigenvalues decrease with increasing dissipation strength, and eventually collapse at the exceptional point $\gamma_c$, indicating the suppression of coherent oscillatory modes. At the same time, the eigenvalues acquire increasingly negative real parts, reflecting enhanced dissipative relaxation. The Liouvillian gap exhibits a pronounced maximum at critical dissipation strength $\gamma_c$, identifying the regime in which relaxation is fastest. For stronger dissipation, the gap decreases, consistent with the onset of a quantum-Zeno-like regime in which strong environmental coupling suppresses coherent population transfer.

An important consequence of our analysis is that the spectral response itself depends on the interaction strength. As $U$ is increased, the maximum of the Liouvillian gap shifts toward larger dissipation strengths, indicating that stronger interactions require stronger environmental coupling to substantially modify the coherent dynamics. This provides a direct connection between the dynamical phase diagram and the underlying relaxation spectrum.

Overall, our results establish self-trapping as a useful dynamical probe of the interplay between coherent many-body evolution and dissipation in a two-mode condensate. The combination of population imbalance and Liouvillian spectral properties gives a unified description of the transition between oscillatory and self-trapping dynamics. These results also show that an engineered dissipation from one of the wells does not merely  damp the coherent motion. Depending on the relative value of the interaction and dissipation strengths, it can destabilize and subsequently restore localization. From a practical angle, this is of great importance, as dissipation can actually be harnessed to create a oscillation-free self-trapped state. Though this state has a finite lifetime because of particle leakage, a non-zero particle number survives within an experimentally relevant time-scale.

Self-trapped dynamics may provide a mechanism for maintaining population imbalance and, under suitable conditions, could be exploited for state preparation and quantum-information applications. Interaction-driven self-trapping in a double well is oscillatory but non-decaying, while dissipation-induced self-trapping is non-oscillatory but has a finite lifetime. These distinct dynamical characteristics offer complementary advantages: interaction-driven self-trapping could be useful for coherence control, precision measurements, and entanglement generation, and dissipation-induced self-trapping may provide a platform for metastable quantum control and the simulation of transient and irreversible dynamics. Combining the two mechanisms could thus provide more versatile control over quantum states and dynamical processes. Importantly, since complete isolation of quantum systems is difficult to achieve experimentally, harnessing dissipation itself to induce self-trapping may offer an additional practical advantage. Because then, an otherwise undesirable and yet unavoidable environmental effect can be converted into a potential control mechanism.

A natural extension of this work would be to investigate the finite-particle-number scaling of the dissipative transition. Since the present study focuses on the qualitative features of the transition, such an analysis lies beyond its scope. Nevertheless, it could provide deeper insights into the statistical properties of the nonequilibrium transition, including its critical behavior and possible universality. Another promising direction would be to explore alternative open-system protocols, such as time-dependent dissipation, dephasing, or particle injection from an external reservoir, and investigate their effects on the resulting quantum dynamics. These extensions could broaden our understanding of self-trapping and, more generally, of nonequilibrium phenomena in open quantum systems. 

.
\section{Acknowledgements}
RD  would like to acknowledge Science and Engineering Research Board (SERB), currently  Anusandhan National Research Foundation (ANRF), Department of Science and Technology, Govt. of India for providing support under the CRG scheme (CRG/2022/007312), and  Rashtriya Uchchatar Shiksha Abhiyan (RUSA) 2.0 (Ministry of Education, Govt. of India).

\bibliography{references}

@article{Gati2007,
doi = {10.1088/0953-4075/40/10/R01},
url = {https://doi.org/10.1088/0953-4075/40/10/R01},
year = {2007},
month = {may},
publisher = {},
volume = {40},
number = {10},
pages = {R61},
author = {Gati, R and Oberthaler, M K},
title = {A bosonic Josephson junction},
journal = {Journal of Physics B: Atomic, Molecular and Optical Physics}
}

@article{Smerzi1997,
	title = {Quantum Coherent Atomic Tunneling between Two Trapped Bose-Einstein Condensates},
	author = {Smerzi, A. and Fantoni, S. and Giovanazzi, S. and Shenoy, S. R.},
	journal = {Phys. Rev. Lett.},
	volume = {79},
	issue = {25},
	pages = {4950--4953},
	numpages = {0},
	year = {1997},
	month = {Dec},
	publisher = {American Physical Society},
	doi = {10.1103/PhysRevLett.79.4950},
	url = {https://link.aps.org/doi/10.1103/PhysRevLett.79.4950}
}

@article{Raghavan1999,
	title = {Coherent oscillations between two weakly coupled Bose-Einstein condensates: Josephson effects, $\ensuremath{\pi}$ oscillations, and macroscopic quantum self-trapping},
	author = {Raghavan, S. and Smerzi, A. and Fantoni, S. and Shenoy, S. R.},
	journal = {Phys. Rev. A},
	volume = {59},
	issue = {1},
	pages = {620--633},
	numpages = {0},
	year = {1999},
	month = {Jan},
	publisher = {American Physical Society},
	doi = {10.1103/PhysRevA.59.620},
	url = {https://link.aps.org/doi/10.1103/PhysRevA.59.620}
}

@article{Albiez2005,
  title = {Direct Observation of Tunneling and Nonlinear Self-Trapping in a Single Bosonic Josephson Junction},
  author = {Albiez, Michael and Gati, Rudolf and F\"olling, Jonas and Hunsmann, Stefan and Cristiani, Matteo and Oberthaler, Markus K.},
  journal = {Phys. Rev. Lett.},
  volume = {95},
  issue = {1},
  pages = {010402},
  numpages = {4},
  year = {2005},
  month = {Jun},
  publisher = {American Physical Society},
  doi = {10.1103/PhysRevLett.95.010402},
  url = {https://link.aps.org/doi/10.1103/PhysRevLett.95.010402}
}

@article{Shchesnovich2010,
	title = {Control of a Bose-Einstein condensate by dissipation: Nonlinear Zeno effect},
	author = {Shchesnovich, V. S. and Konotop, V. V.},
	journal = {Phys. Rev. A},
	volume = {81},
	issue = {5},
	pages = {053611},
	numpages = {5},
	year = {2010},
	month = {May},
	publisher = {American Physical Society},
	doi = {10.1103/PhysRevA.81.053611},
	url = {https://link.aps.org/doi/10.1103/PhysRevA.81.053611}
}

@book{Breuer2007,
	author = {Breuer, Heinz-Peter and Petruccione, Francesco},
	title = {The Theory of Open Quantum Systems},
	publisher = {Oxford University Press},
	year = {2007},
	month = {01},
	isbn = {9780199213900},
	doi = {10.1093/acprof:oso/9780199213900.001.0001},
	url = {https://doi.org/10.1093/acprof:oso/9780199213900.001.0001},
}

@article{Gorini1976,
    author = {Gorini, Vittorio and Kossakowski, Andrzej and Sudarshan, E. C. G.},
    title = {Completely positive dynamical semigroups of N‐level systems},
    journal = {Journal of Mathematical Physics},
    volume = {17},
    number = {5},
    pages = {821-825},
    year = {1976},
    month = {05},
    doi = {10.1063/1.522979},
    url = {https://doi.org/10.1063/1.522979}
}

@Article{Lindblad1976,
author={Lindblad, G.},
title={On the generators of quantum dynamical semigroups},
journal={Communications in Mathematical Physics},
year={1976},
month={Jun},
day={01},
volume={48},
number={2},
pages={119-130},
issn={1432-0916},
doi={10.1007/BF01608499},
url={https://doi.org/10.1007/BF01608499}
}

@article{Julia2010,
  title = {Macroscopic self-trapping in Bose-Einstein condensates: Analysis of a dynamical quantum phase transition},
  author = {Juli\'a-D\'{\i}az, B. and Dagnino, D. and Lewenstein, M. and Martorell, J. and Polls, A.},
  journal = {Phys. Rev. A},
  volume = {81},
  issue = {2},
  pages = {023615},
  numpages = {7},
  year = {2010},
  month = {Feb},
  publisher = {American Physical Society},
  doi = {10.1103/PhysRevA.81.023615},
  url = {https://link.aps.org/doi/10.1103/PhysRevA.81.023615}
}

@article{Milburn1997,
  title = {Quantum dynamics of an atomic Bose-Einstein condensate in a double-well potential},
  author = {Milburn, G. J. and Corney, J. and Wright, E. M. and Walls, D. F.},
  journal = {Phys. Rev. A},
  volume = {55},
  issue = {6},
  pages = {4318--4324},
  numpages = {0},
  year = {1997},
  month = {Jun},
  publisher = {American Physical Society},
  doi = {10.1103/PhysRevA.55.4318},
  url = {https://link.aps.org/doi/10.1103/PhysRevA.55.4318}
}

@article{Buonsante2010,
  title = {Quantum signatures of the self-trapping transition in attractive lattice bosons},
  author = {Buonsante, P. and Penna, V. and Vezzani, A.},
  journal = {Phys. Rev. A},
  volume = {82},
  issue = {4},
  pages = {043615},
  numpages = {9},
  year = {2010},
  month = {Oct},
  publisher = {American Physical Society},
  doi = {10.1103/PhysRevA.82.043615},
  url = {https://link.aps.org/doi/10.1103/PhysRevA.82.043615}
}

@article{Zhou2024,
  title = {Spread complexity and dynamical transition in multimode Bose-Einstein condensates},
  author = {Zhou, Bozhen and Chen, Shu},
  journal = {Phys. Rev. B},
  volume = {110},
  issue = {6},
  pages = {064318},
  numpages = {13},
  year = {2024},
  month = {Aug},
  publisher = {American Physical Society},
  doi = {10.1103/PhysRevB.110.064318},
  url = {https://link.aps.org/doi/10.1103/PhysRevB.110.064318}
}

@article{Soares2025,
title = {Exploring three different indicators of dynamical regimes in a time-dependent two-mode Bose–Hubbard model},
journal = {Physica A: Statistical Mechanics and its Applications},
volume = {671},
pages = {130635},
year = {2025},
issn = {0378-4371},
doi = {https://doi.org/10.1016/j.physa.2025.130635},
url = {https://www.sciencedirect.com/science/article/pii/S0378437125002870},
author = {Gilson Vieira Soares and J.G. Peixoto {de Faria} and Giancarlo Queiroz Pellegrino and Adélcio C. Oliveira},
}

@article{Bardin2026,
  title = {Macroscopic quantum self-trapping in bosonic Josephson junctions: An exact quantum treatment},
  author = {Bardin, A. and Minguzzi, A. and Salasnich, L.},
  journal = {Phys. Rev. A},
  volume = {113},
  issue = {3},
  pages = {L031305},
  numpages = {5},
  year = {2026},
  month = {Mar},
  publisher = {American Physical Society},
  doi = {10.1103/mzg8-gz84},
  url = {https://link.aps.org/doi/10.1103/mzg8-gz84}
}

@article{wuster2025,
  title = {Macroscopic self-trapping in the dynamical tunneling of a Bose-Einstein condensate},
  author = {W\"uster, Sebastian and Cree, Joy and Davis, Matthew J.},
  journal = {Phys. Rev. A},
  volume = {112},
  issue = {6},
  pages = {063305},
  numpages = {16},
  year = {2025},
  month = {Dec},
  publisher = {American Physical Society},
  doi = {10.1103/n8nw-r2hv},
  url = {https://link.aps.org/doi/10.1103/n8nw-r2hv}
}

@article{Secli2021,
doi = {10.1088/1367-2630/ac04c8},
url = {https://doi.org/10.1088/1367-2630/ac04c8},
year = {2021},
month = {jun},
publisher = {IOP Publishing},
volume = {23},
number = {6},
pages = {063056},
author = {Seclì, Matteo and Capone, Massimo and Schirò, Marco},
title = {Signatures of self-trapping in the driven-dissipative Bose–Hubbard dimer},
journal = {New Journal of Physics}
}

@article{Witthaut2008,
  title = {Dissipation Induced Coherence of a Two-Mode Bose-Einstein Condensate},
  author = {Witthaut, D. and Trimborn, F. and Wimberger, S.},
  journal = {Phys. Rev. Lett.},
  volume = {101},
  issue = {20},
  pages = {200402},
  numpages = {4},
  year = {2008},
  month = {Nov},
  publisher = {American Physical Society},
  doi = {10.1103/PhysRevLett.101.200402},
  url = {https://link.aps.org/doi/10.1103/PhysRevLett.101.200402}
}

@article{Witthaut2009,
  title = {Dissipation-induced coherence and stochastic resonance of an open two-mode Bose-Einstein condensate},
  author = {Witthaut, D. and Trimborn, F. and Wimberger, S.},
  journal = {Phys. Rev. A},
  volume = {79},
  issue = {3},
  pages = {033621},
  numpages = {11},
  year = {2009},
  month = {Mar},
  publisher = {American Physical Society},
  doi = {10.1103/PhysRevA.79.033621},
  url = {https://link.aps.org/doi/10.1103/PhysRevA.79.033621}
}

@article{Witthaut2011,
  title = {Beyond mean-field dynamics in open Bose-Hubbard chains},
  author = {Witthaut, D. and Trimborn, F. and Hennig, H. and Kordas, G. and Geisel, T. and Wimberger, S.},
  journal = {Phys. Rev. A},
  volume = {83},
  issue = {6},
  pages = {063608},
  numpages = {7},
  year = {2011},
  month = {Jun},
  publisher = {American Physical Society},
  doi = {10.1103/PhysRevA.83.063608},
  url = {https://link.aps.org/doi/10.1103/PhysRevA.83.063608}
}

@Article{Kordas2015epjs,
author={Kordas, G.
and Witthaut, D.
and Buonsante, P.
and Vezzani, A.
and Burioni, R.
and Karanikas, A. I.
and Wimberger, S.},
title={The dissipative Bose-Hubbard model},
journal={The European Physical Journal Special Topics},
year={2015},
month={Nov},
day={01},
volume={224},
number={11},
pages={2127-2171},
issn={1951-6401},
doi={10.1140/epjst/e2015-02528-2}
}

@article{kordas2015ann,
author = {Kordas, Georgios and Witthaut, Dirk and Wimberger, Sandro},
title = {Non-equilibrium dynamics in dissipative Bose-Hubbard chains},
journal = {Annalen der Physik},
volume = {527},
number = {9-10},
pages = {619-628},
doi = {https://doi.org/10.1002/andp.201400189},
year = {2015}
}

@article{Minganti2018,
  title = {Spectral theory of Liouvillians for dissipative phase transitions},
  author = {Minganti, Fabrizio and Biella, Alberto and Bartolo, Nicola and Ciuti, Cristiano},
  journal = {Phys. Rev. A},
  volume = {98},
  issue = {4},
  pages = {042118},
  numpages = {13},
  year = {2018},
  month = {Oct},
  publisher = {American Physical Society},
  doi = {10.1103/PhysRevA.98.042118},
  url = {https://link.aps.org/doi/10.1103/PhysRevA.98.042118}
}

@article{Vicentini2018,
  title = {Critical slowing down in driven-dissipative Bose-Hubbard lattices},
  author = {Vicentini, Filippo and Minganti, Fabrizio and Rota, Riccardo and Orso, Giuliano and Ciuti, Cristiano},
  journal = {Phys. Rev. A},
  volume = {97},
  issue = {1},
  pages = {013853},
  numpages = {6},
  year = {2018},
  month = {Jan},
  publisher = {American Physical Society},
  doi = {10.1103/PhysRevA.97.013853},
  url = {https://link.aps.org/doi/10.1103/PhysRevA.97.013853}
}

@article{Marko2015,
  title = {Relaxation times of dissipative many-body quantum systems},
  author = {{\ifmmode \check{Z}\else \v{Z}\fi{}}nidari{\ifmmode \check{c}\else \v{c}\fi{}}, Marko},
  journal = {Phys. Rev. E},
  volume = {92},
  issue = {4},
  pages = {042143},
  numpages = {17},
  year = {2015},
  month = {Oct},
  publisher = {American Physical Society},
  doi = {10.1103/PhysRevE.92.042143},
  url = {https://link.aps.org/doi/10.1103/PhysRevE.92.042143}
}

@Article{Fink2018,
author={Fink, Thomas
and Schade, Anne
and H{\"o}fling, Sven
and Schneider, Christian
and Imamoglu, Ata{\c{c}}},
title={Signatures of a dissipative phase transition in photon correlation measurements},
journal={Nature Physics},
year={2018},
month={Apr},
day={01},
volume={14},
number={4},
pages={365-369},
issn={1745-2481},
doi={10.1038/s41567-017-0020-9},
url={https://doi.org/10.1038/s41567-017-0020-9}
}

@article{Kessler2012,
  title = {Dissipative phase transition in a central spin system},
  author = {Kessler, E. M. and Giedke, G. and Imamoglu, A. and Yelin, S. F. and Lukin, M. D. and Cirac, J. I.},
  journal = {Phys. Rev. A},
  volume = {86},
  issue = {1},
  pages = {012116},
  numpages = {21},
  year = {2012},
  month = {Jul},
  publisher = {American Physical Society},
  doi = {10.1103/PhysRevA.86.012116},
  url = {https://link.aps.org/doi/10.1103/PhysRevA.86.012116}
}

@article{Casteels2017,
  title = {Critical dynamical properties of a first-order dissipative phase transition},
  author = {Casteels, W. and Fazio, R. and Ciuti, C.},
  journal = {Phys. Rev. A},
  volume = {95},
  issue = {1},
  pages = {012128},
  numpages = {5},
  year = {2017},
  month = {Jan},
  publisher = {American Physical Society},
  doi = {10.1103/PhysRevA.95.012128},
  url = {https://link.aps.org/doi/10.1103/PhysRevA.95.012128}
}

@article{Mori2023,
  title = {Symmetrized Liouvillian Gap in Markovian Open Quantum Systems},
  author = {Mori, Takashi and Shirai, Tatsuhiko},
  journal = {Phys. Rev. Lett.},
  volume = {130},
  issue = {23},
  pages = {230404},
  numpages = {6},
  year = {2023},
  month = {Jun},
  publisher = {American Physical Society},
  doi = {10.1103/PhysRevLett.130.230404},
  url = {https://link.aps.org/doi/10.1103/PhysRevLett.130.230404}
}

@article{Mori2024,
  title = {Liouvillian-gap analysis of open quantum many-body systems in the weak dissipation limit},
  author = {Mori, Takashi},
  journal = {Phys. Rev. B},
  volume = {109},
  issue = {6},
  pages = {064311},
  numpages = {10},
  year = {2024},
  month = {Feb},
  publisher = {American Physical Society},
  doi = {10.1103/PhysRevB.109.064311},
  url = {https://link.aps.org/doi/10.1103/PhysRevB.109.064311}
}

@article{Mori2023rev,
   author = "Mori, Takashi",
   title = "Floquet States in Open Quantum Systems", 
   journal= "Annual Review of Condensed Matter Physics",
   year = "2023",
   volume = "14",
   number = "Volume 14, 2023",
   pages = "35-56",
   doi = "https://doi.org/10.1146/annurev-conmatphys-040721-015537",
   publisher = "Annual Reviews",
   issn = "1947-5462",
   type = "Journal Article"
  }

@article{Haga2021,
  title = {Liouvillian Skin Effect: Slowing Down of Relaxation Processes without Gap Closing},
  author = {Haga, Taiki and Nakagawa, Masaya and Hamazaki, Ryusuke and Ueda, Masahito},
  journal = {Phys. Rev. Lett.},
  volume = {127},
  issue = {7},
  pages = {070402},
  numpages = {7},
  year = {2021},
  month = {Aug},
  publisher = {American Physical Society},
  doi = {10.1103/PhysRevLett.127.070402},
  url = {https://link.aps.org/doi/10.1103/PhysRevLett.127.070402}
}

@article{Haga2024,
  title = {Oscillating-mode gap: An indicator of phase transitions in open quantum many-body systems},
  author = {Haga, Taiki},
  journal = {Phys. Rev. B},
  volume = {110},
  issue = {10},
  pages = {104303},
  numpages = {15},
  year = {2024},
  month = {Sep},
  publisher = {American Physical Society},
  doi = {10.1103/PhysRevB.110.104303},
  url = {https://link.aps.org/doi/10.1103/PhysRevB.110.104303}
}

@article{Zhu2014,
  title = {Suppressing the Loss of Ultracold Molecules Via the Continuous Quantum Zeno Effect},
  author = {Zhu, B. and Gadway, B. and Foss-Feig, M. and Schachenmayer, J. and Wall, M. L. and Hazzard, K. R. A. and Yan, B. and Moses, S. A. and Covey, J. P. and Jin, D. S. and Ye, J. and Holland, M. and Rey, A. M.},
  journal = {Phys. Rev. Lett.},
  volume = {112},
  issue = {7},
  pages = {070404},
  numpages = {5},
  year = {2014},
  month = {Feb},
  publisher = {American Physical Society},
  doi = {10.1103/PhysRevLett.112.070404},
  url = {https://link.aps.org/doi/10.1103/PhysRevLett.112.070404}
}

@article{Nakagawa2021,
  title = {Exact Liouvillian Spectrum of a One-Dimensional Dissipative Hubbard Model},
  author = {Nakagawa, Masaya and Kawakami, Norio and Ueda, Masahito},
  journal = {Phys. Rev. Lett.},
  volume = {126},
  issue = {11},
  pages = {110404},
  numpages = {6},
  year = {2021},
  month = {Mar},
  publisher = {American Physical Society},
  doi = {10.1103/PhysRevLett.126.110404},
  url = {https://link.aps.org/doi/10.1103/PhysRevLett.126.110404}
}

@article{Nakagawa2020,
  title = {Dynamical Sign Reversal of Magnetic Correlations in Dissipative Hubbard Models},
  author = {Nakagawa, Masaya and Tsuji, Naoto and Kawakami, Norio and Ueda, Masahito},
  journal = {Phys. Rev. Lett.},
  volume = {124},
  issue = {14},
  pages = {147203},
  numpages = {6},
  year = {2020},
  month = {Apr},
  publisher = {American Physical Society},
  doi = {10.1103/PhysRevLett.124.147203},
  url = {https://link.aps.org/doi/10.1103/PhysRevLett.124.147203}
}

@article{Daley2014,
author = {Andrew J. Daley},
title = {Quantum trajectories and open many-body quantum systems},
journal = {Advances in Physics},
volume = {63},
number = {2},
pages = {77--149},
year = {2014},
publisher = {Taylor \& Francis},
doi = {10.1080/00018732.2014.933502},
url = {https://doi.org/10.1080/00018732.2014.933502}
}

@article{Popkov2018,
  title = {Effective quantum Zeno dynamics in dissipative quantum systems},
  author = {Popkov, Vladislav and Essink, Simon and Presilla, Carlo and Sch{\"u}tz, Gunter},
  journal = {Phys. Rev. A},
  volume = {98},
  issue = {5},
  pages = {052110},
  numpages = {8},
  year = {2018},
  month = {Nov},
  publisher = {American Physical Society},
  doi = {10.1103/PhysRevA.98.052110},
  url = {https://link.aps.org/doi/10.1103/PhysRevA.98.052110}
}

@article{Li2026,
doi = {10.1088/1674-1056/adeb60},
url = {https://doi.org/10.1088/1674-1056/adeb60},
year = {2026},
month = {jan},
publisher = {Chinese Physical Society and IOP Publishing Ltd},
volume = {35},
number = {1},
pages = {010306},
author = {Li, Kai-Li and Liu, Yan-Sheng and Zhang, Xi-Zheng},
title = {Control of the Liouvillian gap in the finite open quantum system},
journal = {Chinese Physics B}
}

@article{Streed2006,
  title = {Continuous and Pulsed Quantum Zeno Effect},
  author = {Streed, Erik W. and Mun, Jongchul and Boyd, Micah and Campbell, Gretchen K. and Medley, Patrick and Ketterle, Wolfgang and Pritchard, David E.},
  journal = {Phys. Rev. Lett.},
  volume = {97},
  issue = {26},
  pages = {260402},
  numpages = {4},
  year = {2006},
  month = {Dec},
  publisher = {American Physical Society},
  doi = {10.1103/PhysRevLett.97.260402},
  url = {https://link.aps.org/doi/10.1103/PhysRevLett.97.260402}
}

@article{Barontini2013,
  title = {Controlling the Dynamics of an Open Many-Body Quantum System with Localized Dissipation},
  author = {Barontini, G. and Labouvie, R. and Stubenrauch, F. and Vogler, A. and Guarrera, V. and Ott, H.},
  journal = {Phys. Rev. Lett.},
  volume = {110},
  issue = {3},
  pages = {035302},
  numpages = {5},
  year = {2013},
  month = {Jan},
  publisher = {American Physical Society},
  doi = {10.1103/PhysRevLett.110.035302},
  url = {https://link.aps.org/doi/10.1103/PhysRevLett.110.035302}
}

@Article{Gericke2008,
author={Gericke, Tatjana
and W{\"u}rtz, Peter
and Reitz, Daniel
and Langen, Tim
and Ott, Herwig},
title={High-resolution scanning electron microscopy of an ultracold quantum gas},
journal={Nature Physics},
year={2008},
month={Dec},
day={01},
volume={4},
number={12},
pages={949-953},
issn={1745-2481},
doi={10.1038/nphys1102},
url={https://doi.org/10.1038/nphys1102}
}

@article{
Takafumi2017,
author = {Takafumi Tomita  and Shuta Nakajima  and Ippei Danshita  and Yosuke Takasu  and Yoshiro Takahashi },
title = {Observation of the Mott insulator to superfluid crossover of a driven-dissipative Bose-Hubbard system},
journal = {Science Advances},
volume = {3},
number = {12},
pages = {e1701513},
year = {2017},
doi = {10.1126/sciadv.1701513},
URL = {https://www.science.org/doi/abs/10.1126/sciadv.1701513}
}

@article{Kepesidis2012,
  title = {Bose-Hubbard model with localized particle losses},
  author = {Kepesidis, Kosmas V. and Hartmann, Michael J.},
  journal = {Phys. Rev. A},
  volume = {85},
  issue = {6},
  pages = {063620},
  numpages = {9},
  year = {2012},
  month = {Jun},
  publisher = {American Physical Society},
  doi = {10.1103/PhysRevA.85.063620},
  url = {https://link.aps.org/doi/10.1103/PhysRevA.85.063620}
}

@article{Brazhnyi2009,
  title = {Dissipation-Induced Coherent Structures in Bose-Einstein Condensates},
  author = {Brazhnyi, Valeriy A. and Konotop, Vladimir V. and P{\'e}rez-Garc{\'{\i}}a, V{\'{\i}}ctor M. and Ott, Herwig},
  journal = {Phys. Rev. Lett.},
  volume = {102},
  issue = {14},
  pages = {144101},
  numpages = {4},
  year = {2009},
  month = {Apr},
  publisher = {American Physical Society},
  doi = {10.1103/PhysRevLett.102.144101},
  url = {https://link.aps.org/doi/10.1103/PhysRevLett.102.144101}
}

@article{Labouvie2016,
  title = {Bistability in a Driven-Dissipative Superfluid},
  author = {Labouvie, Ralf and Santra, Bodhaditya and Heun, Simon and Ott, Herwig},
  journal = {Phys. Rev. Lett.},
  volume = {116},
  issue = {23},
  pages = {235302},
  numpages = {5},
  year = {2016},
  month = {Jun},
  publisher = {American Physical Society},
  doi = {10.1103/PhysRevLett.116.235302},
  url = {https://link.aps.org/doi/10.1103/PhysRevLett.116.235302}
}

@Article{Tian2025,
author={Tian, Ye
and Zhao, Yajuan
and Wu, Yue
and Ye, Jilai
and Mei, Shuyao
and Chi, Zhihao
and Tian, Tian
and Wang, Ce
and Shi, Zhe-Yu
and Chen, Yu
and Hu, Jiazhong
and Zhai, Hui
and Chen, Wenlan},
title={Dissipation driven coherent dynamics observed in Bose-Einstein condensates},
journal={Quantum Frontiers},
year={2025},
month={Dec},
day={23},
volume={4},
number={1},
pages={24},
issn={2731-6106},
doi={10.1007/s44214-025-00097-1},
url={https://doi.org/10.1007/s44214-025-00097-1}
}

\end{document}